\documentstyle[twoside,fleqn,epsf,espcrc2]{article}

\newcommand{\AmS}{{\protect\the\textfont2
  A\kern-.1667em\lower.5ex\hbox{M}\kern-.125emS}}

\newcommand{\gsim}{{\protect
  \kern.18em\lower.5ex\hbox{$\stackrel{>}{\sim}$}\kern.25em}}
\newcommand{\lsim}{{\protect
  \kern.17em\lower.5ex\hbox{$\stackrel{<}{\sim}$}\kern.23em}}

\newcommand\be{\begin{equation}}
\newcommand\ee{\end{equation}}
\newcommand\bea{\begin{eqnarray}}
\newcommand\eea{\end{eqnarray}}

\title{
{\normalsize{\vspace*{-3cm} 
\hfill \parbox{30mm}{DESY 96-153}\\[25mm]}}
Low-Lying Eigenvalues of the Wilson-Dirac Operator   
        \thanks{
                Poster presentation at the International
                Symposium on Lattice Field Theory, 4-8 June 1996,
                St. Louis, Mo, USA}}

\author{K. Jansen$^{\rm a}$, C.Liu \address{Deutsches Elektronen Synchroton, 
        DESY, Notkestr. 85, 22603 Hamburg, Germany}, %
        H. Simma$^{\rm b}$ and D. Smith\address{Dept. of Physics and Astronomy,
        The University of Edinburgh, Edinburgh EH9 3JZ, Scotland \\
        ~~(UKQCD Collaboration)}}
           
\begin{document}

\begin{abstract}
An exploratory study of the low-lying eigenvalues  
of the 
Wilson-Dirac operator and their corresponding eigenvectors is presented. 
Results for the eigenvalues from quenched and unquenched simulations 
are discussed. 
The eigenvectors are studied with respect to their localization properties
in the quenched approximation for the cases of SU(2) and SU(3). 
\end{abstract}

\maketitle

\section{Introduction}

In this contribution we consider the eigenvalues
and corresponding eigenvectors of the 
{\em hermitean} operator
\be \label{q}
Q=\gamma_5 M/\left(1+8\kappa\right)\; ,
\ee
where $M$ is the Wilson-Dirac operator
defined through
\bea
 [M\psi](x) & = & \psi(x) \nonumber\\
  & - & \kappa \sum_{\mu=1}^4 
   \Bigl\{(1-\gamma_{\mu} )\,U_{\mu}(x)\,\psi(x+\mu) \Bigr.
  \nonumber\\
  & + & \Bigl.(1+\gamma_{\mu})\,U^{\dagger}_{\mu}(x-\mu)\,\psi(x-\mu) 
  \Bigr\}\ .
\label{Dirac}
\eea
The hopping parameter $\kappa$ is related to the
bare quark mass $m_0$ by $\kappa=(8+2m_0)^{-1}$.
The gauge fields $U_{\mu}(x)$ are mostly taken to be in
SU(2) and for some cases in SU(3).
The matrix $Q$ is normalized in such a way that the eigenvalues
$\lambda$ of $Q$ satisfy $0\le \lambda^2 \le 1$.
In the following we will concentrate on those eigenvalues of $Q$
closest to zero, and refer to this part of the spectrum as the
``low-lying'' eigenvalues.

The eigenvalues of $Q$ are of crucial importance for understanding
the performance of algorithms used in dynamical fermion simulations. 
In particular, the scaling behaviour of the algorithms when approaching 
the chiral limit can be expressed in terms of the lowest eigenvalue 
of $Q^2$ \cite{karl}. 

Furthermore, the eigenmodes themselves play an interesting
physical r\^ole. In this work we will only address the space-time
localization properties of the eigenvectors belonging to 
low-lying eigenvalues.
Additional more challenging questions include
their connection to chiral symmetry breaking and topology.
Another important aspect is the contribution of the low-lying 
eigenvalues to physical observables like the pion or $\rho$ meson 
correlation functions. 

We compute the low-lying eigenvalues and the corresponding eigenvectors 
by using a modified Conjugate Gradient algorithm to minimize the 
Ritz functional  
\be \label{ritz}
\frac{<\Psi_k |Q^2 | \Psi_k>}{<\Psi_k |\Psi_k>}\; .
\ee
For the computation of higher eigenvalues the minimization of (\ref{ritz})
is restricted to $\Psi_k$ being orthogonal to previously calculated 
eigenvectors.
Convergence is accelerated by intermediate diagonalizations in the space 
spanned by $\{\Psi_i\}$ \cite{simma}.
This method has the virtue of being numerically stable and 
of providing rigorous error bounds for the eigenvalues calculated.
Moreover the approximate eigenvectors are obtained without additional cost.

\section{Low-lying eigenvalues for SU(2)} 

We first investigate how the spectrum changes with the gauge coupling 
$g^2_0$. Recall that for free Wilson fermions and $\kappa < 1/8$, 
the lowest eigenvalue of $Q^2$ is \emph{8-fold} degenerate.
Working in the quenched approximation on an $8^4$ lattice at 
$\kappa = 0.12$, we find that up to $g^2_0 = 4/\beta$ around $0.8$ 
the lowest \emph{eight} eigenvalues are
\emph{nearly degenerate} and well separated from the ninth lowest.
Above $g^2_0 \approx 0.8$ the eigenvalues form a dense band. 
It is interesting to note that this value of $\beta$, where the gap to 
the ninth eigenvalue closes, is roughly where one might expect the 
finite temperature phase transition. 

\begin{figure} \label{figure1}  
\centerline{ \epsfysize=11.0cm 
             \epsfxsize=8.0cm 
             \epsfbox{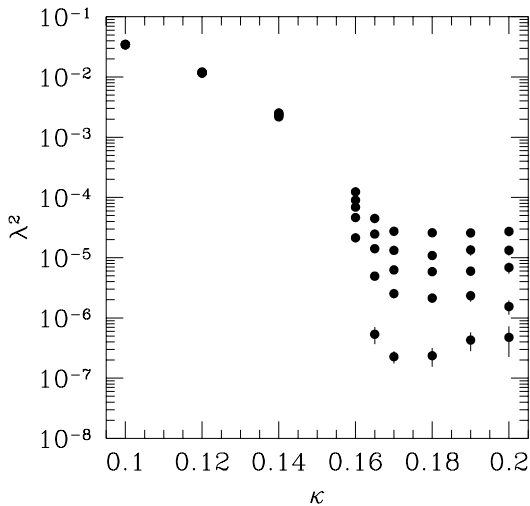}}
\vspace{-10mm}
\caption{Low-lying eigenvalues as a function of $\kappa$ for quenched
SU(2) configurations at $\beta=2.3$.
}
\vspace{-0mm}
\end{figure}

The $\kappa$ dependence of the $n$ lowest eigenvalues (with $n=1,3,5,7,10$) 
at $\beta=2.3$ is shown in Fig.~1 for a lattice
size of $12^4$. The results in Fig.~1 are averaged over 20 
configurations each separated by 500 pure gauge updates.
For $\kappa < 0.16$ the low-lying eigenvalues
are almost degenerate and appear on top of each other in the graph. 
For $\kappa \ge 0.16$ the eigenvalues drop significantly and  
their density becomes smaller.

The coarse $\kappa$-scan in Fig.~1 provides only an overall 
picture of the behaviour of the eigenvalues. On individual
configurations, used for Fig.~1, a much finer $\kappa$-scan,
guided by a perturbative expansion in small changes in $\kappa$,
reveals almost-zero modes with values of $\lambda^2$ that can drop 
to e.g. $O(10^{-18})$, many orders of magnitude smaller than the lowest 
eigenvalue in Fig.~1. 
So far, it seems that these almost-zero modes appear {\em above} 
$\kappa_c$. If this observation is true for all
coupling parameter values of interest, these almost-zero modes would
be harmless. Otherwise they would indicate the breakdown of the
quenched approximation since any occurance of such a mode could
severely distort the configuration sample. 

\begin{figure} \label{figure2}  
\centerline{ \epsfysize=11.0cm 
             \epsfxsize=8.0cm 
             \epsfbox{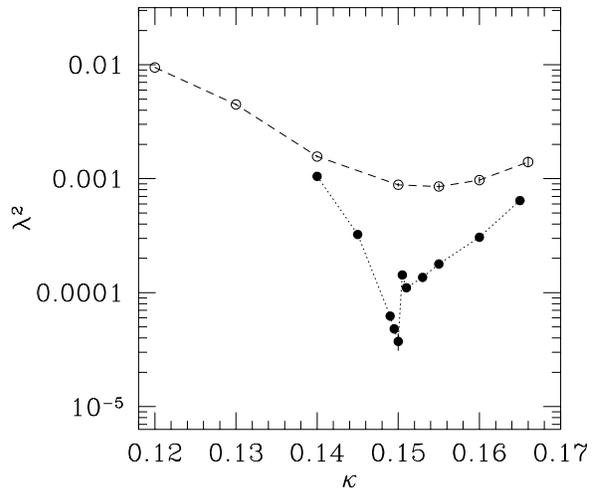}}
\vspace{-10mm}
\caption{ Lowest eigenvalue on $8^4$ (open symbols) and
          $16^4$ (full symbols) 
          lattices from unquenched simulations 
          as a function of $\kappa_{\rm sea}$.
         }
\vspace{-0mm}
\end{figure}

%
To investigate the situation for dynamical fermions we made  unquenched runs 
for SU(2) at $\beta =2.3$.
In Fig.~2 we show the lowest eigenvalue on $8^4$ and $16^4$ lattices.
For the $16^4$ lattice, the scan in the sea-$\kappa$ 
has a high resolution. On some of these configurations
we also performed fine valence-$\kappa$ scans.
In both cases we did not find any
indication of the existence of the almost-zero modes. 
Fig.~2 also shows that for the $16^4$ lattice the lowest eigenvalue 
in unquenched simulations has a sharp minimum. This suggests that 
the lowest eigenvalue could be a good quantity to identify the
phase transition.

\section{Localization of the eigenmodes}
\begin{figure} \label{figure3}  
\centerline{ \epsfysize=11.0cm 
             \epsfxsize=8.0cm 
             \epsfbox{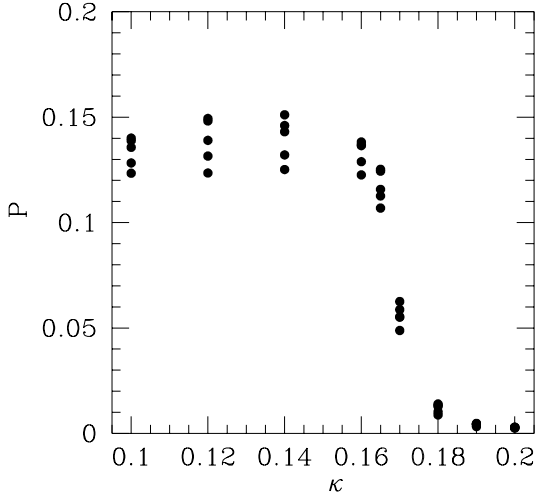}}
\vspace{-10mm}
\caption{Participation ratio for the eigenvalues in Fig.~1.}
\end{figure}

A simple quantity to study localization properties of 
wave functions is the participation ratio \cite{parti} 
\be \label{p}
P=\frac{1}{V}\frac{(\sum_{x,\alpha,a} \vert\Psi_{\alpha a}(x)\vert^2)^2}
                   {\sum_{x,\alpha,a} \vert\Psi_{\alpha a}(x)\vert^4},
\ee
where $V$ is the lattice volume and $(\alpha,a)$ are Dirac and colour indices.
For pointlike localization one obtains $P\rightarrow 0$,
whereas $P\rightarrow 1$ for a uniform distribution.
The participation ratios which correspond to the eigenvalues of Fig.~1
are shown in Fig.~\ref{figure3}. They clearly indicate strongly localized 
eigenmodes for $\kappa > 0.17$, but do not give a clear indication 
at lower $\kappa$ values.

\begin{figure} \label{figure4}
\vspace{-11.5mm}
\centerline{ \epsfysize=11.0cm 
             \epsfxsize=7.0cm 
             \epsfbox{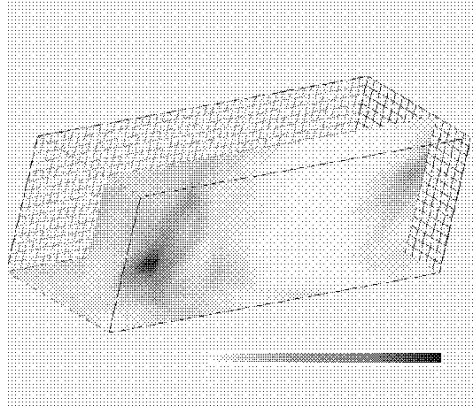}}
\vspace{-15mm}
\caption{ 3-dimensional slices of $\rho(x)$ at $\kappa = 0.1475$.
   The grey scale from bright to dark corresponds to values of the density
   $\rho(x)$ between 0 and $\rho_{\rm max}$.
 }
\vspace{-5mm}
\end{figure}
\begin{figure} \label{figure5}
\centerline{ \epsfysize=11.0cm 
             \epsfxsize=7.0cm 
             \epsfbox{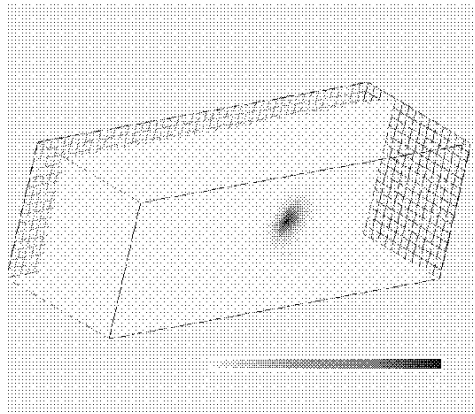}}
\vspace{-15mm}
\caption{ Same as Fig.~4, but at $\kappa = 0.165$. }
\vspace{-0mm}
\end{figure}

To investigate this question further we plot the local 
density $\rho(x)= \|\Psi(x)\|^2$ for 3-dimensional cuts through a lattice
of size $16^3 48$ for quenched SU(3) at $\beta=6.0$.
Fig.~4 indicates localization also for 
$\kappa < \kappa_c \approx 0.157$;
however it is less pronounced than for $\kappa > \kappa_c$ (Fig.~5).

For a quantitative analysis of the localization we define
\be \label{rv}
R_V(c) = \frac{1}{V}\#\left\{x: \rho(x)\ge c\cdot\rho_{\rm max}\right\}
\ee
which measures the relative number of lattice sites with a 
local density larger than a certain fraction $c$ ($0\le c \le 1$)
of the maximum value $\rho_{\rm max}$. We find that practically
all sites contributing to $R_V(c)$ are in a connected region, and 
that $R_V(c)$ scales proportional to the inverse lattice volume.

To characterize the localization we try for
$\rho(x)$ a parametrization
$\rho(x) \sim \exp\left\{-m_{loc}\| x\|\right\}$, 
where $\| x\|$ denotes the distance from the 
site where $\rho=\rho_{\rm max}$. The effective ``localization
mass'' $m_{loc}$ is the local rate with which $\rho(x)$ decays. 
Taking for $\| x\|$ the effective radius
$r_{\rm eff}(c) = \sqrt[4]{2 R_V(c)/\pi^2}$ one obtains 
$m_{loc}=-\ln(c)/r_{\rm eff}$.
A plateau in the c-dependence of $m_{loc}$ would then indicate that $\rho(x)$ 
decays exponentially according to the above ansatz.

\vspace{-0mm}
\begin{figure}[htb] \label{figure6}  
\centerline{ \epsfysize=7.0cm 
             \epsfxsize=7.0cm 
             \epsfbox{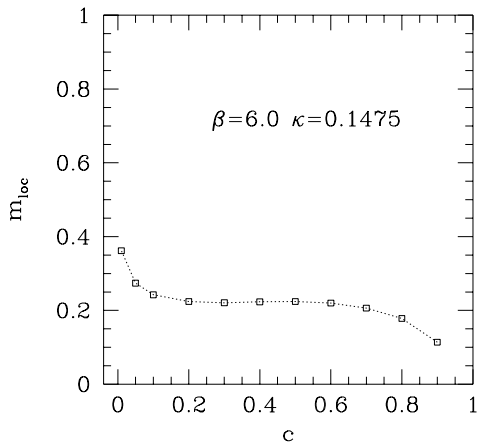}}
\vspace{-10mm}
\caption{Localization mass on a $16^348$ lattice from quenched simulations 
   in SU(3).}
\vspace{-0mm}
\end{figure}

This is indeed observed as Fig.~6 shows for a value of
$\kappa$ well below $\kappa_c$. 
One finds a nice stable plateau in $m_{loc}$ completely consistent 
with an exponential localization.

\section{Conclusion} 

We have presented an exploratory study of the low-lying eigenvalue 
spectrum of the hermitian Wilson-Dirac operator
$Q$ given in eq.(\ref{q}). 
In the quenched approximation we find that the lowest eigenvalues drop 
sharply when the critical hopping parameter is reached. 
The isolated almost-zero modes we find in the quenched approximation
at $\beta=2.3$ 
have not occurred in our unquenched runs.
The eigenmodes corresponding to the low-lying eigenvalues show 
an exponential localization for $\kappa < \kappa_c$. For
$\kappa$ above $\kappa_c$ they are even stronger (almost pointlike) 
localized. 
In general we did not see any qualitative difference between the results 
from SU(2) and SU(3).

Although the eigenvalues of $Q$ can not be directly related to the 
complex spectrum of $M$, we expect 
the qualitative behaviour of the eigenmodes to be similar. 
Of course, this relation has to be explored.
It will also be interesting to relate the low-lying eigenvalues to
topological properties of the theory and to quantify
the contribution of the low-lying eigenvalues to physical observables.
We hope to return to these questions in future publications.

\section*{Acknowledgement}
We like to thank M.~B\"aker and M.~L\"uscher for helpful discussions and
suggestions. The numerical work was carried out on APE/Quadrics and T3D
systems and we thank DESY-IfH (Zeuthen) and The University of Edinburgh 
for providing the necessary resources. We acknowledge financial support 
by PPARC grant GR/J21347 and by the Carnegie Trust for The Universities 
of Scotland.

\def\sp{$\;$}

\end{document}